\documentclass[aps,prb,twocolumn,floatfix,floats,superscriptaddress]{revtex4}
\usepackage{graphicx,latexsym}
\usepackage{amsmath}
\begin{document}

\title{Transport through a quantum ring, a dot and a barrier \\ embedded in a
       nanowire in magnetic field}

\author{Vidar Gudmundsson}
\affiliation{Science Institute, University of Iceland,
        Dunhaga 3, IS-107 Reykjavik, Iceland}
\author{Yu-Yu Lin}
\affiliation{Department of Electrophysics, National Chiao Tung
        University, Hsinchu 30010, Taiwan}
\author{Chi-Shung Tang}
\affiliation{Physics Division, National Center for Theoretical
        Sciences, P.O.\ Box 2-131, Hsinchu 30013, Taiwan}
\author{Valeriu Moldoveanu}
\affiliation{Institute of Physics and Technology of Materials,
             P.O.\ Box MG7, Bucharest-Magurele, Romania}
\author{Jens Hjorleifur Bardarson}
\affiliation{Science Institute, University of Iceland,
        Dunhaga 3, IS-107 Reykjavik, Iceland}
\author{Andrei Manolescu}
\affiliation{Science Institute, University of Iceland,
             Dunhaga 3, IS-107 Reykjavik, Iceland}
\affiliation{Institute of Physics and Technology of Materials,
             P.O.\ Box MG7, Bucharest-Magurele, Romania}

%

\begin{abstract}
We investigate the transport through a quantum ring, a dot and a
barrier embedded in a nanowire in a homogeneous perpendicular
magnetic field. To be able to treat scattering potentials of finite
extent in magnetic field we use a mixed momentum-coordinate
representation to obtain an integral equation for the multiband
scattering matrix. For a large embedded quantum ring we are able to
obtain Aharanov-Bohm type of oscillations with superimposed narrow
resonances caused by interaction with quasi-bound states in the
ring. We also employ scattering matrix approach to calculate the
conductance through a semi-extended  barrier or well in the wire.
The numerical implementations we resort to in order to describe the
cases of weak and intermediate magnetic field allow us to produce
high resolution maps of the ``near field'' scattering wave
functions, which are used to shed light on the underlying scattering
processes.
\end{abstract}

\pacs{78.67.-n,75.75.+a,72.30.+q}

\maketitle
\def\mb#1{\mbox{\boldmath$#1$}}
\def\t#1{\text{#1}}
\def\nn{\nonumber \\}

\def\sec#1{Sec.~\ref{#1}}
\def\eq#1{Eq.~(\ref{#1})}
\def\fig#1{Fig.~\ref{#1}}

%
\section{Introduction}

The influence of a single impurity on the conductance of a quasi 
one-dimensional quantum channel has been investigated by several
groups theoretically\cite{Chu89:5941,Bagwell90:10354,Vargiamidis02:302,Cattapan03:903} 
and experimentally.\cite{Faist90:3217} Commonly the impurities are
considered to be short range and represented by a $\delta$-function, 
though treatments of more extended scatterers, like square 
barriers,\cite{Bagwell90:10354} can be found. Recently, 
the application to nanosized systems, has 
spurred the use of general methods built on the Lippmann-Schwinger
equation or the equivalent $T$-matrix formalism to describe the 
scattering of more general extended potentials in 
quantum channels\cite{Bardarson04:01}
or curved wires.\cite{Qu04:085414}   

The inclusion of a constant homogeneous magnetic field perpendicular 
to the quasi one-dimensional electron channel or wire drastically changes the
properties of the system. Without the magnetic field a centered symmetric
scattering potential leads to ``selection rules'' that restrict 
the possible scattering processes. These are lifted by the magnetic field
resulting in a rich structure contrasted to the conductance steps in an 
ideal wire as long as the magnetic length is not much shorter than the 
width of the wire and the range of the scattering potential.\cite{Takagaki95:7017}     

The character of the Lorentz force does not allow us to
establish a simple multimode formulation of the scattering process
in configuration space,\cite{Cattapan03:903} but in strong magnetic field
Gurvitz used a scheme to develop a multimode formalism using a 
Fourier Transform with respect to the transport direction,
and a truncation to a two-mode formalism allowed him to
seek analytical solutions for a short range scatterer
present in a wire with general confinement.\cite{Gurvitz95:7123}

Here we extend this formalism by noting that in the case of a
parabolic shape of the wire confinement we obtain
coupled Lippmann-Schwinger equations with a nonlocal scattering potential
in Fourier space for the different modes. A transformation of this
system of equations to corresponding equations for the $T$-matrix shows that
it bears a strong resemblance to the corresponding equations for the
system in no external magnetic field.\cite{Bardarson04:01} 
We exploit this fact to seek numerical solutions for the system     
in weak and intermediate strength of the magnetic field where a 
two mode approximation is not always warranted. One benefit of the
numerical approach is that it allows us to map out with high resolution the
probability density for the scattering states near the scatterer. 
These ``near field'' solutions give us a good indication of the
scattering process itself. We explore a quantum wire with an
embedded quantum dot or a ring. We are able to increase the size of
the ring to the limit that we observe Aharanov-Bohm type of oscillations.

In order to investigate the scattering of potentials that are homogeneous in the direction
perpendicular to the wire representing a barrier or a well  
we employ an alternative method based on mode matching. The smooth scattering potential is 
sliced into a series of $\delta$-potentials. The scattering matrix is
then constructed from repeated mode matching at each slice. 

%
\section{Models}

We consider electron transport along a parabolically confined quantum wire 
parallel to the $x$-axis and perpendicular to a homogeneous magnetic 
field ${\bf B} = B{\bf \hat z}$. In the center of the wire the electrons
are scattered by a potential $V_{sc}({\bf r})$ to be specified below. 
The system under investigation is described by
the Hamiltonian $H = H_0 + V_{\rm sc}(\bf{r})$ with
\begin{equation}
      H_0 = \frac{\hbar^2}{2m^*}\left[ -i  \mb{\nabla} - \frac{eB}{\hbar
      c} y \mb{\hat x} \right]^2 + V_{\rm c}(y),
\label{Hamiltonian}
\end{equation}
where the wire is assumed to be parabolically confined, namely
$V_{\rm c}(y)$ = $m^* \Omega_0^2 y^2 / 2$ with $m^*$ and $\Omega_0$
being, respectively, the effective mass of an electron in GaAs-based
material and the confinement parameter. Here $-e$ is the charge of
an electron. We present two quite different approaches to describe
the scattering process of incoming electron states, one using
wave function matching appropriate to describe scattering from a 
thin homogeneous barrier or well perpendicular the wire, 
and the other one using a $T$-matrix formalism
to calculate the scattering of an embedded quantum dot or ring in 
the wire.

%
\subsection{Scattering matrix approach}

\begin{figure}[thb]
      \includegraphics[width=0.38\textwidth]{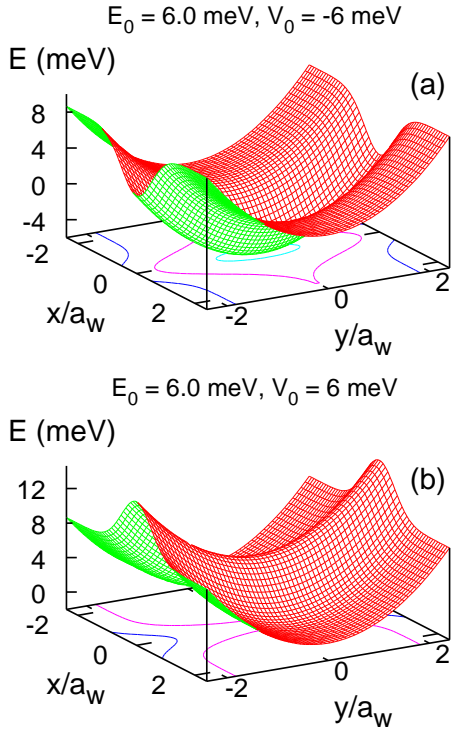}
      \caption{Barrier or well embedded in a quantum wire with
      $V_0$ = (a) $-6$ meV and (b) $6$ meV, respectively.  Other
      parameters are $B=0$ T, $a_w=33.7$ nm, $E_0=\hbar\Omega_0=6.0$ meV,
      and $\beta a_w^2=1.897$.} \label{barrier_well}
\end{figure}

In this section, we employ the scattering matrix approach to the
calculation of coherent electronic transport in a quantum wire in
the presence of a Gaussian-profile scattering potential.  The
potential under investigation can be realized as a finger gate atop
the wire and can be modeled by
\begin{equation} \label{GP1}
      V_{\rm sc}({\bf r})= V_{0}\exp \left( -\beta x^2 \right),
\end{equation}
as is shown in \fig{barrier_well}.

To obtain a dimensionless expression we employ the Bohr radius
$a_0$, and have thus the relevant units for the confinement
parameter ${\Omega_0}^* = \hbar / (m^*{a_0}^2)$, and the magnetic
field $B^*$ = $\hbar c / (e {a_0}^2)$. As such, by defining the unit
of the cyclotron frequency to be $\omega_c^*$ = $\Omega_0^*$, the
cyclotron frequency simply has the dimensionless form $\omega_c$ =
$B$.

In the absence of scatterers the eigenfunctions of $H_0$ can be
written as~\cite{Gurvitz95:7123}
\begin{equation}
      \psi^{\pm}(x,y,k_n) = e^{\pm ik_n x}\chi^{\pm}(y,k_n)
\end{equation}
with $\pm k_n$ being the wave vector along $\pm \mb{\hat x}$ in the
$n$th transverse subband, where $\chi^{\pm}(y,k_n)$ satisfies the
reduced dimensionless equation
\begin{eqnarray}
      \left[ -\frac{\partial^2}{\partial y^2} +
      \Omega_w^2 \left( y \mp \alpha_{n} \right)^2 \right]\chi^{\pm}(y,k_n)
      = E_{n}\chi^{\pm}(y,k_n),
\end{eqnarray}
which is a harmonic oscillator of frequency $\Omega_w$ with a
shifted center $\alpha_{n}$ = $\omega_c k_n/\Omega_w^2$. These
eigenmodes of the electron in a state described by $\psi(x,y,k_n)$ in a pure
quantum wire have the energy spectrum
\begin{equation}
      E = E_n + {\cal K}(k_n),
\end{equation}
composed of Landau levels $E_n$ = $\left( n+1/2 \right)\Omega_w$, where $\Omega_w$
= $\sqrt{\omega_c^2+\Omega_0^2}$, shifted by the
confining potential, and the kinetic energy  ${\cal K}(k_n)$
= $k_n^2 (\Omega_0/\Omega_w)^2$. The eigenfunctions
$\chi^{\pm}(y,k_n)$ of the eigenmodes are given by
\begin{equation}
      \chi^{\pm}(y,k_n) = N_n H_n\left[ \sqrt{\Omega_w} \left( y \mp
      \alpha_n \right) \right] \exp \left[ -\frac{\Omega_w}{2}  \left( y
      \mp \alpha_n \right)^2 \right],
\end{equation}
where $N_n = \left(
\Omega_w /\pi \right)^{1/4} \left( 2^n n! \right)^{-1/2}$ is
a normalization constant.

To utilize the scattering matrix approach, we divide the Gaussian
scattering potential into a series of slices of width $\delta L$,
each of them is described by a delta-profile potential
\begin{equation}
      V_{\rm sc}\left( x_i \right) =V_{0}\delta L \exp \left( -\beta x_i^2
      \right) \delta\left(x-x_i \right).
\end{equation}
The Gaussian potential Eq.\ (\ref{GP1}) can thus be described by
$V_{\rm sc}\left({\bf r}\right) = \sum_{i=1}^{N_L} V_{\rm
sc}\left(x_i \right)$ if we divide the potential into sufficiently
many pieces.  For a right-going incident wave $\psi(x,y,k_n)$ from
the $n$th mode of the left reservoir, the corresponding scattering
wave function can be expressed in the form
\begin{eqnarray}
      \psi_n^{(i)}(x,y,k_n) &=& e^{ik_n x} \chi_n^{+}\left(y,k_n\right)
      \nonumber \\
      && + \sum\limits_{n'} { r^{i}_{n'n} e^{-ik_{n'}x} \chi _{n'}^{-}
      \left(y,k_{n'}\right) }
\end{eqnarray}
for $x < x_i$, and
\begin{equation}
      \psi_n^{(i)}(x,y,k_n) = \sum\limits_{n'} { t^{i}_{n'n} e^{-ik_{n'}x} \chi _{n'}^{+}
      \left(y,k_{n'}\right) }
\end{equation}
for $x > x_i$.  Following similar procedure we can also obtain the
reflection and transmission coefficients, $\tilde{r}^{i}_{n'n}$ and
$\tilde{t}^{i}_{n'n}$, for a left-going incident wave.

The scattering should satisfy two boundary conditions, one
requirement is that $\psi(x,y)$ has to be continuous at $x = x_i$
and the other one stems from integration of the Schr{\"o}dinger equation
across $x = x_i$. Multiplying these boundary conditions by the
eigenfunction $\chi_m(y)$ of the ordinary unshifted harmonic oscillator with
confining frequency $\Omega_w$, and integrating over $y$, we obtain
\begin{equation}\label{ME1}
      \sum\limits_{n'}{I_{mn'}^{+}\left( k_{n'}\right) \, t^{i}_{n'n} } -
      \sum\limits_{n'}{I_{mn'}^{-}\left( k_{n'}\right) \,r^{i}_{n'n} } =
      I_{mn}^{+}\left( k_n \right)
\end{equation}
\begin{equation}\label{ME2}
      \sum\limits_{n'}{J_{mn'}^{+}\left( k_{n'}\right) \, t^{i}_{n'n} } +
      \sum\limits_{n'}{K_{mn'}^{-}\left( k_{n'}\right) \,r^{i}_{n'n} } =
      k_n I_{mn}^{+}\left( k_n \right)
\end{equation}
where the matrix elements are related to the overlap integrals
\begin{equation}
      I_{mn'}^{\pm} = \int \chi_m(y) \chi_{n'}^{\pm}(y) dy \, ,
\end{equation}
\begin{equation}
      J_{mn'}^{+}\left( k_{n'}\right) = k_{n'} I_{mn'}^{+} \left( k_{n'}
      \right) + i V_0 V_{mn'}^{+}
\end{equation}
with
\begin{equation}
      V_{mn'}^{+} = \int \chi_m \left( y \right) V_{\rm sc}(y)
      \chi_{n'}^{\pm}(y) dy\, ,
\end{equation}
and
\begin{equation}
      K_{mn'}^{-}\left( k_{n'} \right) = k_{n'} I_{mn'}^{-}\left( k_{n'}
      \right)\, .
\end{equation}

Using Eqs.\ (\ref{ME1})-(\ref{ME2}) and the corresponding equations
for $\tilde{r}^{i}_{n'n}$ and $\tilde{t}^{i}_{n'n}$, one can
establish the scattering matrix for the Gaussian-shape
potential.~\cite{Tang00:127} From the ratio of the transmitted
and the incident current we obtain the current
transmission $T_{\beta\alpha}$, in which $\alpha$ and $\beta$ are,
respectively, the incident and the transmitting lead. Since the
current flowing into the terminal $R$ (drain) from the system should
take into account the current into the terminal $L$ (source) from
the system, the net charge current into terminal $R$ is therefore
given by
\begin{eqnarray}
      I(V) &=&  \frac{2e}{h}\int\limits_0^{\mu + eV/2} dE\,
      f_L(E) T_{RL}(E) \nonumber \\
      && - \frac{2e}{h} \int\limits_0^{\mu - eV/2} dE\, f_R(E) T_{LR}(E)
\end{eqnarray}
where $f(E)$ indicates the Fermi distribution function and $\mu$ is
the zero-bias Fermi level.  The two-terminal conductance is given by
$G = \partial I/\partial V$. Below we assume that the scattering
potential is located at the center of the nanowire, the applied
perpendicular magnetic field is homogeneous, and the source-drain
bias is sufficiently low. Hence, the zero temperature conductance
can be expressed in terms of the incident electron energy $E$ in the
simple form
\begin{equation} \label{GE}
      G(E) = \frac{2e^2}{h} \sum\limits_{n=0}^{N} T_n(E)\, ,
\end{equation}
where $N$ denotes the highest propagating mode incident from the
source electrode.  The current transmission coefficient $T_n(E)$ for
an electron incident in the $n$th subband from the source electrode
is given by
\begin{equation}
      T_{n}(E) = \sum_{n'(\mu_{n'}>0)} \frac{k_{n'}}{k_n} \left|
      t_{n'n}^{RL} \right|^2 .
\end{equation}
The current reflection coefficient $R_n(E)$ can be calculated by a
similar form to get current conservation condition for checking the
numerical accuracy.

%
\subsection{T-matrix formalism}

\begin{figure}[thb]
      \includegraphics[width=0.48\textwidth]{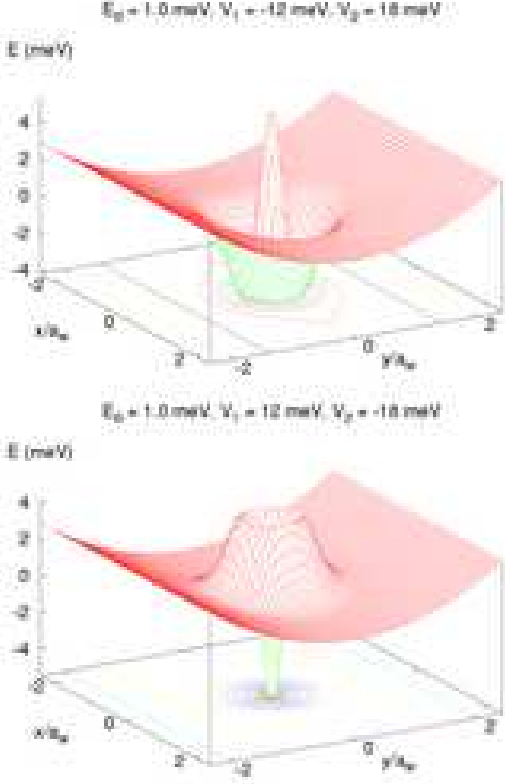}
      \caption{Quantum ring (upper) or dot (lower) embedded in a quantum wire,
               $B=0$ T, $a_w=33.7$ nm, $E_0=\hbar\Omega_0=1.0$ meV,
               $\beta_1a_w^2=3.41$, and $\beta_2a_w^2=11.37$.}
\label{Ring_dot}
\end{figure}

We consider a quantum dot or a ring embedded in the wire 
and parameterize the scattering potential accordingly
combining two Gaussian functions of different shapes 
\begin{equation}
      V_{sc}({\bf r}) = V_1\exp{(-\beta_1 r^2)}+V_2\exp{(-\beta_2 r^2)}, 
\label{Vsc}
\end{equation}
as is shown in Fig.\ \ref{Ring_dot}.
Together the magnetic field and the parabolic confinement define a natural
length scale $a_w=\sqrt{\hbar/(m^*)\Omega_w}$, where 
$\Omega_w=\sqrt{\omega_c^2+\Omega_0^2}$, with the cyclotron frequency
$\omega_c=eB/(m^*c)$, is the natural frequency of the quantum wire in a
magnetic field.

Along the lines of Gurvitz we choose to use the mixed momentum-coordinate 
presentation of the wave functions\cite{Gurvitz95:7123}
\begin{equation}
      \Psi_E(p,y)=\int dx\:\psi_E(x,y)e^{-ipx},
\label{Psi_py}
\end{equation}
and expand them in channel modes
\begin{equation}
      \Psi_E(p,y)=\sum_n\varphi_n(p)\phi_n(p,y),
\label{Psi_py_exp}
\end{equation}
i.e.\ in terms of the eigenfunctions for the pure parabolically confined wire in
magnetic field 
\begin{equation}
      \phi_n(p,y)=\frac{\exp{[-\frac{1}{2}(\frac{y-y_0}{a_w})^2]}}
      {\sqrt{2^n\sqrt{\pi}n!a_w}}H_n\left(\frac{y-y_0}{a_w}\right),
\label{Phi_para}
\end{equation}
with the center coordinate $y_0=pa_w^2\omega_c/\Omega_w$. 
These eigenmodes of the pure quantum wire have the energy spectrum
$E_{np} = E_n + \kappa_n (p)$ with $E_n =\hbar\Omega_w(n+1/2)$ and
\begin{equation}
      \kappa_n(p)=\frac{(pa_w)^2}{2}\frac{(\hbar\Omega_0)^2}{\hbar\Omega_w}.
\label{Kappa}
\end{equation}  
Using Eq.'s (\ref{Psi_py}-\ref{Kappa}) and performing a Fourier transform with
respect to the coordinate $x$ transforms the Schr{\"o}dinger equation 
corresponding to the Hamiltonian (\ref{Hamiltonian}) into a system of 
coupled nonlocal integral equations 
in momentum space
\begin{equation}
      \kappa_n(q)\varphi_n(q) +
      \sum_{n'}\int\frac{dp}{2\pi}\: V_{nn'}(q,p)\varphi_{n'}(p)
      =(E-E_n)\varphi_n(q),
\label{Nonleq}
\end{equation}  
where 
\begin{equation}
      V_{nn'}(q,p) = \int dy\: \phi_n^*(q,y)V(q-p,y)\phi_{n'}(p,y),
\label{Vpq}
\end{equation}  
and 
\begin{equation}
      V(q-p,y) = \int dx\: e^{-i(q-p)x}V_{sc}(x,y).
\label{Vp-qy}
\end{equation}  
The matrix elements (\ref{Vpq}-\ref{Vp-qy}) for the scattering potential
(\ref{Vsc}) can be evaluated analytically since they consist of Gaussians
and Hermite polynomials, see Appendix \ref{sec:appendix}.

The special form of the part of the energy dispersion $\kappa_n(q)$ for parabolic
confinement allows us now to rewrite Eq.\ (\ref{Nonleq}) as
\begin{equation}
     \begin{split}
      \left[ -(qa_w)^2+(k_n(E)a_w)^2\right] &\varphi_{n}(q)\\
      =\frac{2\hbar\Omega_w}{(\hbar\Omega_0)^2}\sum_{n'}
      &\int\frac{dp}{2\pi}\: V_{nn'}(q,p)\varphi_{n'}(p),
      \end{split}
\label{NonEq}
\end{equation}  
where we have defined the effective band momentum $k_n(E)$ as
\begin{equation}
      (E-E_n)=\frac{[k_n(E)]^2}{2}\frac{(\hbar\Omega_0)^2}{\hbar\Omega_w}.
\label{kn(E)}
\end{equation}  
In the absence of a magnetic field it is possible to derive an equivalent
effective one-dimensional multi-band Schr{\"o}dinger equation to (\ref{NonEq})
in coordination space.\cite{Cattapan03:903} 
This multi-band equation is then usually
transformed into a system of effective one-dimensional coupled Lippmann-Schwinger integral
equations that is convenient for numerical computation. 
Here we can proceed along these lines, but the magnetic field forces us to 
do this in momentum space where we shall
see that the corresponding Lippmann-Schwinger equations are better transformed
into integral equations for the $T$-matrix in order
to facilitate numerical evaluation.
Considering Eq.\ (\ref{NonEq}) it is clear that the incoming scattering states
satisfy
\begin{equation}
      \left[ -(qa_w)^2+(k_n(E)a_w)^2\right]\varphi_{n}^0(q)=0,
\label{G-1phy}
\end{equation}  
which implies a Greens function
\begin{equation}
      \left[ -(qa_w)^2+(k_n(E)a_w)^2\right]G_E^{n}(q)=1.
\label{Geq}
\end{equation}  
The Greens function can now be used to write down coupled Lippmann-Schwinger
equations in momentum space
\begin{equation}
      \varphi_n(q) = \varphi_n^0(q) + G_E^n(q)\sum_{n'}\int\frac{dpa_w}{2\pi}
      {\tilde V}_{nn'}(q,p)\varphi_{n'}(p),
\label{LSeq}
\end{equation} 
where ${\tilde V}_{nn'}(q,p)=V_{nn'}(q,p)2\hbar\Omega_w/[a_w(\hbar\Omega_0)^2]$.
These equations are inconvenient for numerical evaluation as the in-state 
$\varphi_n^0$ is proportional to a Dirac delta function. Symbolically Eq.\
(\ref{LSeq}) can be expressed as $\varphi = \varphi^0 + G{\tilde V}\varphi$, and an
iteration of the equation gives $\varphi = \varphi^0 + G{\tilde V}\varphi^0 +
G{\tilde V}G{\tilde V}\varphi^0 + \cdots = (1+G{\tilde T})\varphi^0$, where
we have introduced the $T$-matrix satisfying the
symbolic equation ${\tilde T} = {\tilde V} + {\tilde V}G{\tilde T}$. 
Fully written the equation determining the $T$-matrix is
\begin{equation}
      \begin{split}
      {\tilde T}_{nn'}(q,p) =& {\tilde V}_{nn'}(q,p)\\ +& 
      \sum_{m'}\int\frac{dka_w}{2\pi}
      {\tilde V}_{nm'}(q,k)G_E^{m'}(k){\tilde T}_{nm'}(k,p).
      \end{split}
\label{Teq}
\end{equation} 
This set of equations is easier to solve numerically than the equivalent Lippmann-Schwinger
equations (\ref{LSeq}) after the singularities of the Greens function have
been handled with special care.\cite{JHB:04} We obtain analytically the
contribution of the poles of the Greens function and perform the remaining
principal part integration by removing the singularity by a subtraction
of a zero.\cite{Haftel1:70,Landau:96}  

Comparison with the nonseparable two-dimensional Lippmann-Schwinger equation in
configuration space for extended scattering potential in a magnetic field
gives the connection between the $T$-matrix and the probability amplitude for transmission
in mode $n$ with momentum $k_n$ if the in-state is in mode $m$ with momentum
$k_m$
\begin{equation}
      t_{nm}(E) = \delta_{nm} - \frac{i\sqrt{(k_m/k_n)}}{2(k_ma_w)}
      \left(\frac{\hbar\Omega_0}{\hbar\Omega_w} \right)^2{\tilde T}_{nm}(k_n,k_m).
\label{t(E)}
\end{equation} 
The conductance is then according to the Landauer-B{\"u}ttiker formalism
defined as
\begin{equation}
      G(E) = \frac{2e^2}{h}{\rm Tr}[ {\bf t}^{\dagger}(E){\bf t}(E)] ,
\label{G}
\end{equation} 
where ${\bf t}$ is evaluated at the Fermi energy. 

Symbolically the wave function can be expressed as $\varphi = (1+G{\tilde T})\varphi^0$ 
if the in-state $\varphi^0$ is given. Together with Eq.\ (\ref{Psi_py}) and 
(\ref{Psi_py_exp}) this gives
\begin{equation}
      \begin{split}
            \psi_E(x,y) =&e^{ik_nx}\phi_n(k_n,y) \\
            +&\sum_m\int\frac{dqa_w}{2\pi}e^{iqx}G_E^m(q){\tilde T}_{mn}(q,k_n)
            \phi_m(q,y) ,
      \end{split}
\label{PsiE_xy}
\end{equation} 
for an incident electron with energy $E$, in mode $n$ with momentum $k_n$.
To calculate the wave function the same methods are used to isolate the
contribution from the poles of the Greens function as were used for the
calculation of ${\tilde T}$ with Eq.\ (\ref{Teq}).

%
\section{Results}
\subsection{Embedded barrier}

\begin{figure}[tbhp!]
      \includegraphics[width=0.6\textwidth, angle=270]{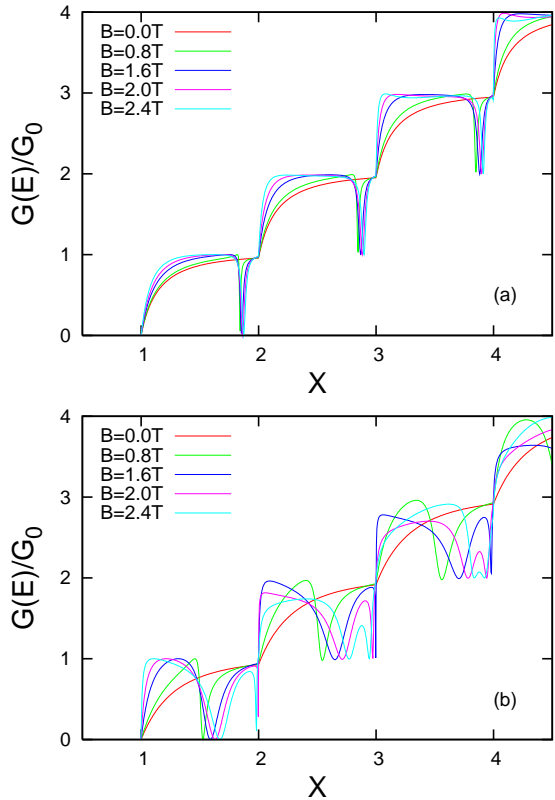}
      \caption{Conductance of a parabolically confined wire as a function
      of the energy parameter $X = E/\hbar\Omega_w + 1/2$ 
      for various applied perpendicular magnetic fields. The
      amplitudes of the attractive impurity potential are (a) $-6$ meV and
      (b) $-12$ meV. Other parameters are $\hbar\omega_0 = 6$ meV and
      $\beta_2 a_w^2(B=0)=1.897$.} \label{VmBx}
\end{figure}
In this section, we present our numerical results of exploring
electronic transport properties using a Gaussian-shape potential
model described by \eq{GP1} -- the conductance versus the incident
electron energy $E$.  The parameters used to obtain our numerical results
are taken from the GaAs-$\t{Al}_\t{x}\t{Ga}_\t{1-x}\t{As}$
heterostructure system.  The values that we choose for our
material parameters are $E_{\text{Ryd}} = 5.93$ meV and  $a_0 =
9.79$ nm.

The conductance \eq{GE} of the wire is presented in \fig{VmBx} for
several values of magnetic field. To explore the transport
properties it is convenient to show the conductance as a function of
energy of the incoming electron state scaled by the subband energy
level spacing $X = E/\hbar\Omega_w + 1/2$ such that the integral
values of $X$ indicates the number of incident modes. In \fig{VmBx},
we present the conductance for magnetic fields with strengths from
$0$ to $2.4$ T for either weak ($V_0=-6$ meV) or strong ($V_0=-12$
meV) attractive potentials, as shown in \fig{VmBx}(a) and (b),
respectively. For the case of weak attractive potential shown in
\fig{VmBx}(a), one can see that the dip structures in $G(E)$ are
pinned at around $X \simeq n+0.85$, and the location
is insensitive to the magnetic field.  It turns out that these
structures correspond to the electrons incident from subband $n$
scattered elastically into the $n+1$ subband threshold forming
quasi-bound states.~\cite{Bardarson04:01}  It can be demonstrated
that these quasi-bound states are formed in the leads out of the
embedded Gaussian potential.~\cite{Tang03:205324}

For the case of strong attractive potential shown in \fig{VmBx}(b),
one can see that there are two types of quasi-bound state
features.  The mechanism of sharp dips below the subband threshold
is similar to the case of weak attractive potential.  On the other
hand, it is interesting to see the valley structures in \fig{VmBx}
for $B\ne 0$. These valleys correspond to quasi-bound states
formed in the attractive Gaussian potential.  When the applied
magnetic field becomes stronger, the blue shift of these valleys
indicates such quasi-bound states are formed closer to the edge of
the Gaussian potential due to the cyclotron motion. The large width
of these valley structures implies the short life time of these
quasi-bound states. When increasing the strength of the magnetic
field, these valleys become wider.  This indicates that the
electrons in high magnetic field with short cyclotron radius easily
escape from the quasi-bound states formed in such a strong
attractive potential.  We note in passing that in the absence of
magnetic field, the intersubband transition is forbidden
since the attractive potential is uniform in the transverse
direction, and we cannot see any dip structures in $G(E)$.

\begin{figure}[tbhp!]
      \includegraphics[width=0.3\textwidth, angle=270]{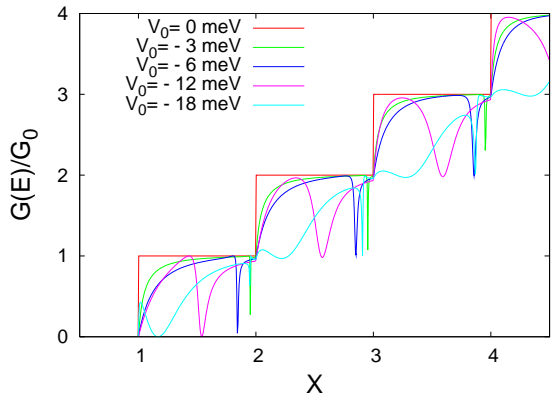}
      \caption{Conductance of a parabolically confined wire as a function
      of incident electron energy for various amplitudes of attractive
      potential. The other parameters are taken to be $B=1$ T,
      $\hbar\omega_0 = 6$ meV, and $\beta_2 a_w^2(B=0)=1.897$. }
\label{VmxB1d0}
\end{figure}
In Fig.~\ref{VmxB1d0}, we study how the conductance can be
affected by changing the amplitude of the attractive potential by
fixing the strength of the magnetic field $B = 1$ T, the
confining potential $\hbar \omega_y = 6$ meV, and the Gaussian
parameter $\beta_2 a_w^2(B=0)$ = $1.897$.  In the
absence of the Gaussian potential (red curve), the electron
transport manifests an ideal quantized conductance, the magnetic
field plays no role. When increasing the amplitude of the attractive
potential, the subband levels in the potential will
decrease in energy. Therefore, we can find a red shift
of the quasi-bound states. More precisely, for the cases of $V_0$ =
$-3,\,-6,\,-12$, and $-18$ meV, the dip structures occur at around
$E/\hbar\Omega_w$ = $1.95$, $1.84$, $1.54$, and $1.17$,
respectively, in the attractive potential. It is also
interesting to note that when the attractive potential is very
strong, such as $V_0 = -18$ meV, one can see a second dip structure
appearing below the subband thresholds, both are quasi-bound states of
the attractive potential.

\begin{figure}[tbhp!]
      \includegraphics[width=0.6\textwidth, angle=270]{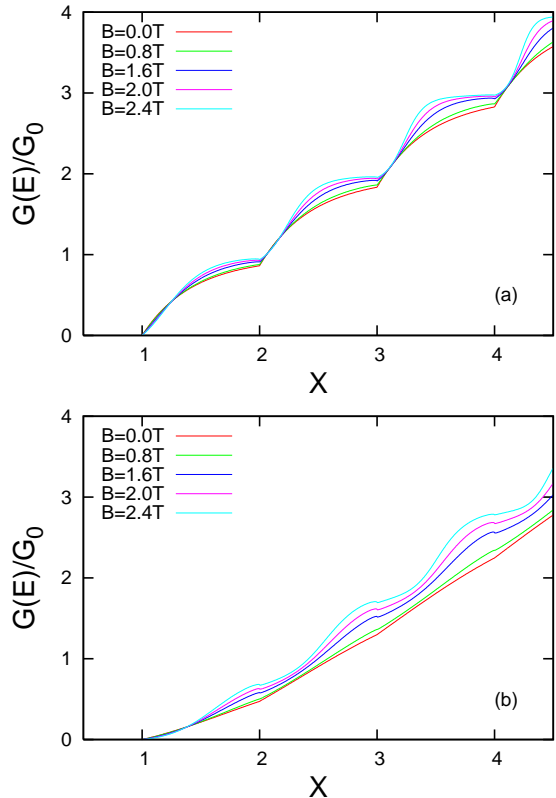}
      \caption{Conductance  as a function of incident electron energy
      with various applied magnetic fields.
      The amplitudes of the repulsive potential barrier are
      $V_0$ = (a) $6$ meV and (b) $12$ meV. Other parameters are $\hbar\omega_0$ = $6$ meV.
      and $\beta_2 a_w^2(B=0)$ = $1.897$.}
\label{VpBx}
\end{figure}
Figure \ref{VpBx} shows the conductance as a function of incident
electron energy for several values of magnetic field in the presence
of a repulsive potential. The magnetic fields are tuned from $0$ to
$2.4$ T for either weak ($V_0=6$ meV) or strong ($V_0=12$ meV)
repulsive potentials, as shown in \fig{VpBx}(a) and (b),
respectively. For the case of weak repulsive potential shown in
\fig{VpBx}(a), one can see that the conductance plateaus are
suppressed from the ideal case. When increasing the applied
perpendicular magnetic field, the suppressed conductance plateaus
tend to be enhanced back to the ideal case.
For the case of strong
repulsive potential (see \fig{VpBx}(b)), the conductance curves are
suppressed much more then those for the weak repulsive one. Without a
magnetic field, the conductance behaves according to Ohm's law.
When the applied magnetic field is increased, the conductance can be
enhanced, and the quantization behavior becomes slowly recognizable.

\begin{figure}[tbhp!]
      \includegraphics[width=0.3\textwidth, angle=270]{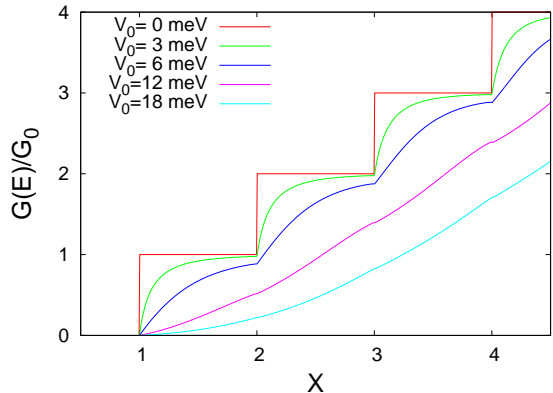}
      \caption{Conductance as a function of incident electron energy with
      various amplitudes of repulsive potential. Other parameters are
      taken to be $B=1$ T, $\hbar\omega_0 = 6$ meV, and $\beta_2
      a_w^2(B=0)=1.897$. } \label{VpxB1d0}
\end{figure}
One may want to see how the conductance can be affected by
changing the amplitude of the repulsive potential, as is shown in
\fig{VpxB1d0}. We fix the strength of the magnetic field $B = 1$ T,
the confining potential $\hbar \omega_y = 6$ meV of the nanowire,
and the potential configuration $\beta_2 a_w^2(B=0)$ = $1.897$. In
the absence of scattering potential, the conductance is ideally
quantized. When increasing the amplitude of the repulsive potential,
the electrons are reflected by the barrier and the conductance is
strongly suppressed. For weak barrier, such as $V_0$ = $3$ and $6$
meV, the quantization behavior in $G(E)$ is still visible. However,
when the barrier is strong enough, such as $V_0$ = $12$, and $18$
meV, not only the conductance is strongly suppressed, but the
quantization behavior in $G(E)$ has vanished.

To conclude this section, we note in passing that when the
scattering potential (well or barrier) is uniform in the transverse
direction it does not break the translational invariance along the
lateral confining direction. However, in the presence of magnetic
field, if such a scattering potential is a well then one can find
quasi-bound states due to elastic intersubband transitions to a
higher subband threshold. However, if the scattering potential is a
barrier, one finds no quasi-bound state features even in a magnetic
field up to 2.4 T.

%
\subsection{Embedded quantum ring and dot}

\begin{figure}[tbhp!]
      \includegraphics[width=0.45\textwidth]{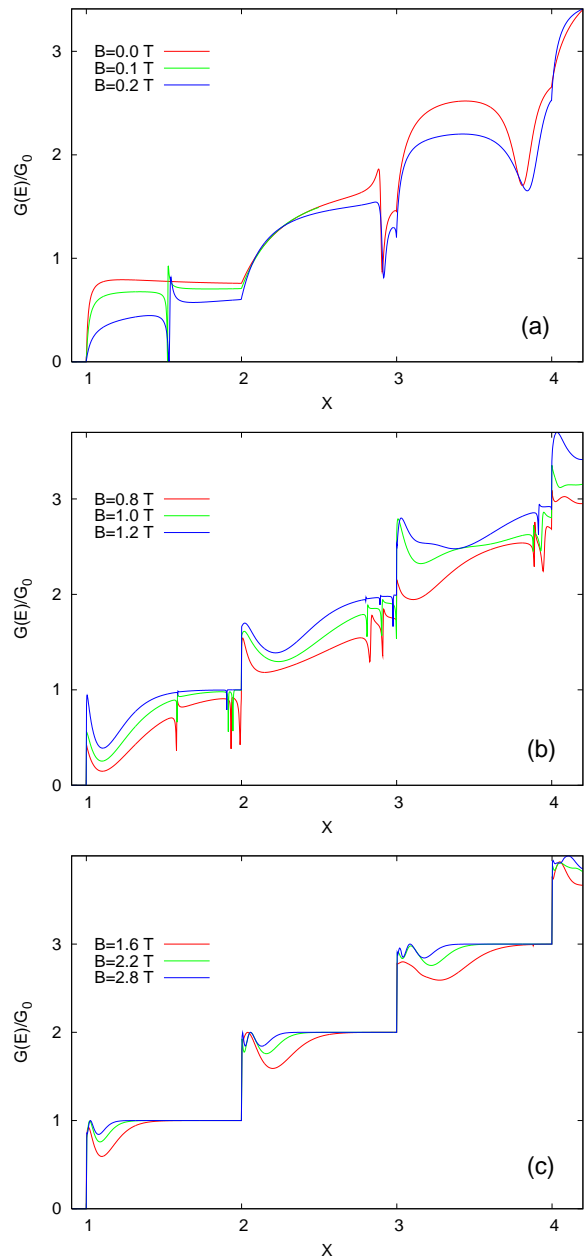}
      \caption{The conductance of a parabolic wire with an embedded ring 
               in units of $G_0=2e^2/h$. $E_w=\hbar\Omega_w$, $V_1=-12$ meV, $V_2=18$ meV,
               $\hbar\Omega_0=1.0$ meV, $\beta_1 a_w^2(B=0)=3.412$, 
               $\beta_2 a_w^2(B=0)=11.37$, and 9 subbands are included.}
\label{fig:C_dhd}
\end{figure}

To model an embedded quantum ring with the parameterization
(\ref{Vsc}) we initially choose the parameters used in Fig.\ \ref{Ring_dot},
such that when $B=0$ then $\beta_1a_w^2=3.41$, $\beta_2a_w^2=11.37$,
and $\hbar\Omega_w = 1.0$ meV. (The parameters of the potential (\ref{Vsc}),
$\beta_1$ and $\beta_2$ do not depend on $B$, but $a_w$ does). 
$V_1= -12$ meV, and $V_2=18$ meV.
We are thus investigating relatively broad wire with a small embedded ring
structure with diameter of approximately 40 nm. We assume the wire
to be a GaAs wire as mentioned above. The conductance (\ref{G}) of the wire
is presented in Fig.\ \ref{fig:C_dhd} for several values of the magnetic field.
To compare the results for various values of the magnetic field it is 
convenient to observe the conductance as function of the energy of the
incoming electron state scaled by the distance of the energy subbands,
i.e.\ $E/(\hbar\Omega_w)=E/E_w$ and furthermore use 
$X = E/\hbar\Omega_w + 1/2$ such that an integral
value of $X$ indicates the number of incident modes.

\begin{figure}[tbhp!]
      \includegraphics[width=0.45\textwidth]{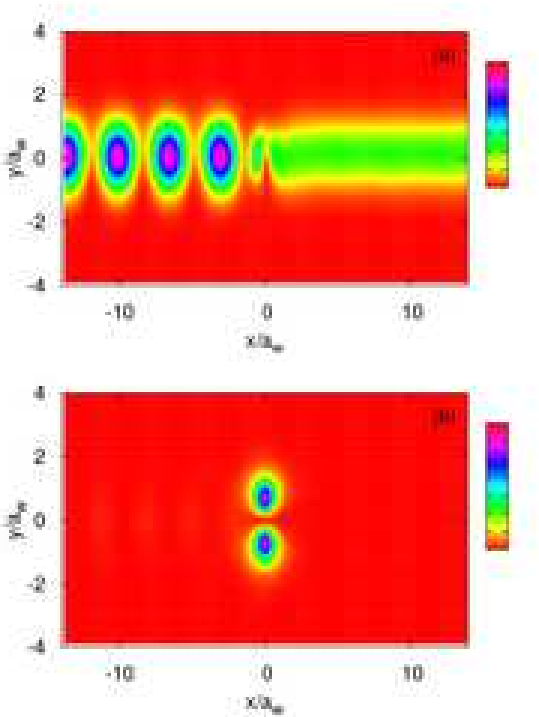}
      \caption{The probability density of the scattering state $\psi_{E}(x,y)$
               in the parabolic quantum wire in the presence of an
               embedded quantum ring (Fig.~\ref{Ring_dot}), corresponding to
               the conductance in Fig.\ \ref{fig:C_dhd}a at $B=0.1$ T.               
               The incident energy $X=1.4$ (a), and
               $X=1.538$ (b) corresponding to the dip
               in the conductance.}  
\label{fig:Wf.B01_dhd_1}
\end{figure}

In Fig.\ \ref{fig:C_dhd}a we see that as soon as the magnetic field is different
from zero a strong Fano-like\cite{Fano61:1866,Simpson63:158} 
resonance dip appears in the first plateau just 
above $X = 1.5$. As we argue below the dip corresponds to
a destructive quantum interference between a quasi-bound state
in the ring and an in-state of the wire. 
Figure \ref{fig:Wf.B01_dhd_1} displays the
total probability to find an electron in the wire close to the
scattering center, the quantum ring. The probability is calculated
using the wave function (\ref{PsiE_xy}) for two values of the energy of
the incoming electron in the lowest transverse mode, $n=0$. 
Just below the resonance at $X = 1.4$ 
Fig.\ \ref{fig:Wf.B01_dhd_1}a reveals us a normal scattering process.
The scattering only takes place very close to $x\approx 0$ and on the 
left hand side we see the interference pattern for the incoming and
the reflected wave. On the right hand side the electrons only travel in one 
transverse mode and only to the right so we have a constant 
probability already a short distance away from the scattering center. 
The situation is quite different in Fig.\ \ref{fig:Wf.B01_dhd_1}b that
displays the probability density for the state exactly in the resonance dip.
Here no transmitted wave is present, but the probability close to the 
quantum ring is high enough that the probability for the incoming and
the reflected waves is not visible on the color scale used. 

The symmetry of the quasi-bound state indicates that it is    
an evanescent state belonging to the second subband $n=1$.
Without a magnetic field the scattering via the evanescent
state in the second subband is forbidden in the case of a 
symmetric potential placed in the middle of the wire.   
In that case a dip occurs in the second band due to a scattering
through a evanescent state in the third 
subband.\cite{Bagwell90:10354,Gurvitz93:10578,Bardarson04:01}

\begin{figure}[tbhp!]
      \includegraphics[width=0.45\textwidth]{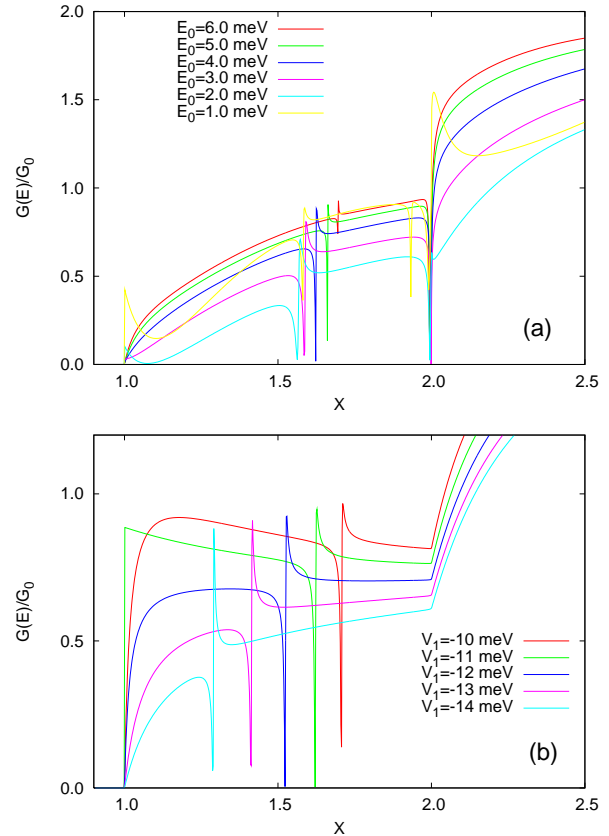}
      \caption{The conductance of a parabolic wire with an embedded ring 
               in units of $G_0=2e^2/h$ as a function of the wire confinement
               $E_0=\hbar\Omega_0$ (a), and depth of the ring $V_1$ (b).
               $V_2=18$ meV, $\beta_1 a_w^2(B=0)=3.412$, 
               $\beta_2 a_w^2(B=0)=11.37$, and 9 subbands are included.
               $E_w=\hbar\Omega_w$, $B=0.8$ T and $V_1=-12$ meV in (a). 
               $E_0=1.0$ meV and $B=0.1$ T in (b).}
\label{fig:C_dhd_Fano}
\end{figure}

In order to further support our view that the resonance is due to a
quasi-bound state of the quantum ring located in the continuum of the first
subband (The ring lowers a state in the second subband into the first one),
we see in Fig.\ \ref{fig:C_dhd_Fano}a how the broadening or narrowing of the 
wire has little effects on the energy of the state. On the other hand,
as seen in Fig.\ \ref{fig:C_dhd_Fano}b the energy of the quasi-bound state
changes linearly with the depth of the ring potential.  

\begin{figure}[tbhp!]
      \includegraphics[width=0.45\textwidth]{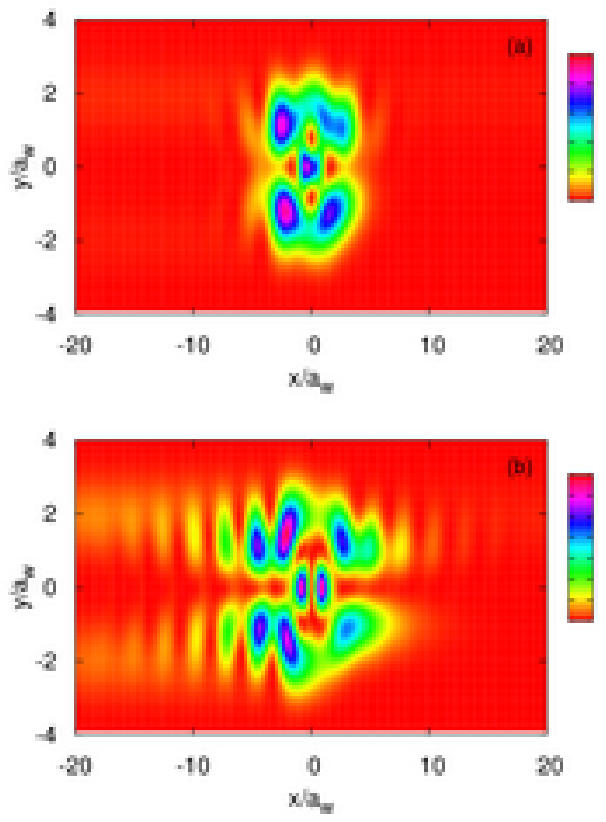}
      \caption{The probability density of the scattering state $\psi_{E}(x,y)$
               in the parabolic quantum wire in the presence of an
               embedded quantum ring (Fig.~\ref{Ring_dot}), corresponding to
               the conductance in Fig.\ \ref{fig:C_dhd}b at $B=0.8$ T.               
               The incident energy $X=1.933$ (a), and
               $X=1.991$ (b), corresponding to two minima
               in the conductance at the end of the first step.}
\label{fig:Wf.B01_dhd_2}
\end{figure}

For some intermediate values of the magnetic field we see other minima
occurring in the conductance closer to the end of the first step.
For example, for $B=0.8$ T this is visible in Fig.\ \ref{fig:C_dhd}b
at $X = 1.933$ and $1.991$. The corresponding probability
densities are seen in Fig.\ \ref{fig:Wf.B01_dhd_2}
The symmetry of both densities indicates that the dips are caused by 
scattering via evanescent states of the second subband just as the
dip in the middle of the first conductance step. These states are
quasi-bound states of the ring further in the continuum of the
first subband. The higher state, Fig.\ \ref{fig:Wf.B01_dhd_2}b
has acquired more of the character of the geometry of the wire than
the ring, and it extends far beyond the ring.
The presence of Fano lineshapes in the conductance is not surprising
as the mesoscopic Fano effect was already experimentally reported for both
a single electron transistor\cite{Gores00:2188} and an Aharanov-Bohm
interferometer with an embedded quantum 
dot.\cite{Kobayashi02:256806,Kobayashi03:235304}
Nevertheless, in these two experiments the wire was much smaller than the 
mesoscopic system that caused the Fano interference. The results
presented here suggest that this may be observed also in the case
of a broad wire. 

At still higher magnetic field, see Fig.\ \ref{fig:C_dhd}c, the
conductance has approached the ideal case as the magnetic field
has now squeezed the wave functions together and closer to the
edge as soon as the momentum is different from zero. The wave function
thus bypasses the scattering potential. We shall see this effect
clearer below.      

\begin{figure}[tbhp!]
      \includegraphics[width=0.45\textwidth]{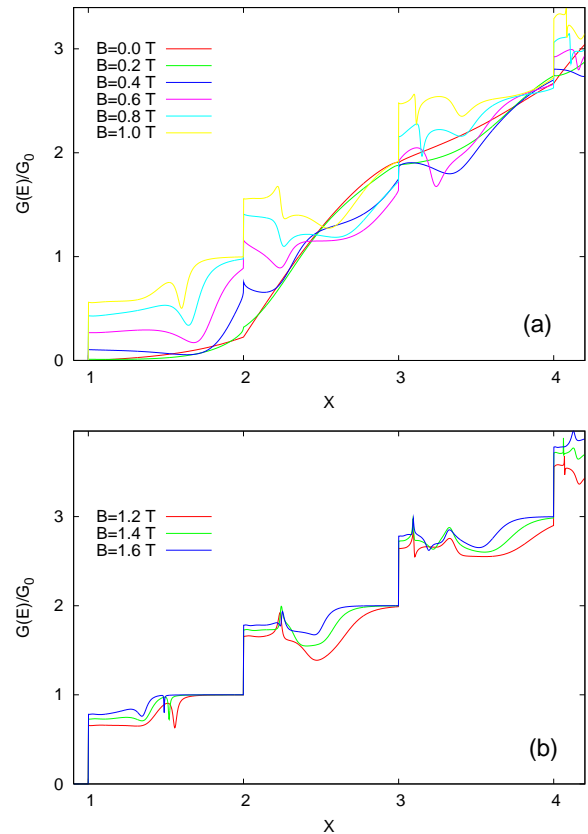}
      \caption{The conductance of a parabolic wire with an embedded dot
               in units of $G_0=2e^2/h$. $E_w=\hbar\Omega_w$, 
               $V_1=12$ meV, $V_2=-18$ meV,
               $\hbar\Omega_0=1.0$ meV, $\beta_1 a_w^2(B=0)=3.412$, 
               $\beta_2 a_w^2(B=0)=11.37$, and 9 subbands are included.}
\label{fig:C_hdh}
\end{figure}

\begin{figure}[tbhp!]
      \includegraphics[width=0.45\textwidth]{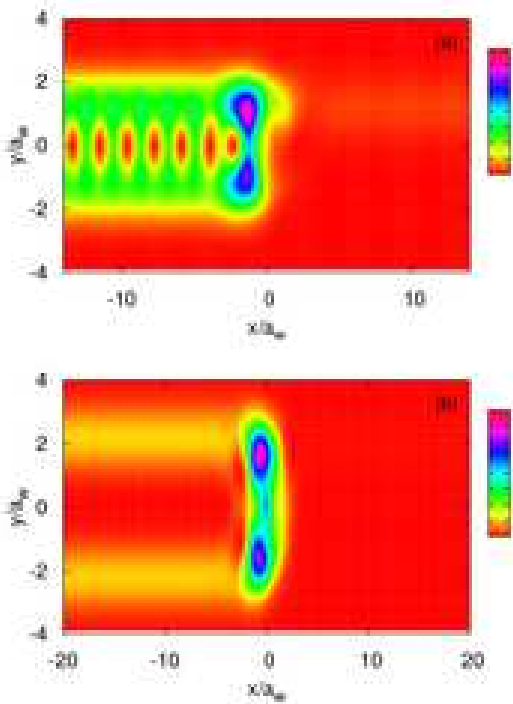}
      \caption{The probability density of the scattering state $\psi_{E}(x,y)$
               in the parabolic quantum wire in the presence of an
               embedded quantum dot (Fig.~\ref{Ring_dot}), corresponding to
               the conductance in Fig.\ \ref{fig:C_hdh}.               
               The incident energy $X = 1.679$ at $B=0.6$ T (a), and
               $X = 1.557$ at $B=1.2$ T (b), corresponding to two minima
               in the conductance at the end of the first step.}
\label{fig:B06_12_hdh_1}
\end{figure}

The conductance of a wire with an embedded dot is presented in
Fig.\ \ref{fig:C_hdh}.
The effects of an increasing magnetic field become very clear if we
compare the probability density for the dips at $X = 1.679$ 
when $B=0.6$ T, and the one at $X = 1.557$ when $B=1.2$ T,
see Fig.\ \ref{fig:B06_12_hdh_1}.
Both cases show a partial blocking of the channel due to backscattering 
caused by a quasi-bound state created by an evanescent state of the second
subband, but the main difference is the total separation of the incoming
and the reflected channel at the higher magnetic field. At the lower
magnetic field we still see a interference pattern between these channels.

\begin{figure}[tbhp!]
      \includegraphics[width=0.45\textwidth]{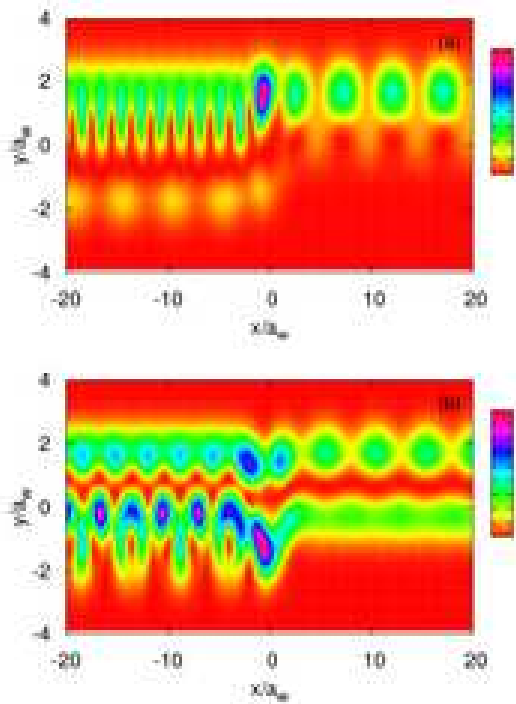}
      \caption{The probability density of the scattering state $\psi_{E}(x,y)$
               in the parabolic quantum wire in the presence of an
               embedded quantum dot (Fig.~\ref{Ring_dot}), corresponding to
               the conductance in Fig.\ \ref{fig:C_hdh}a at $B=0.6$ T
               for a state with incident energy $X = 2.235$,               
               in mode $n=0$ (a), and $n=1$ (b).}
\label{fig:B06_hdh_n12_3}
\end{figure}

To explore further the scattering from the embedded dot when there is not
a complete separation between the edge and bulk channels we show the
probability density for $X = 2.235$ at $B=0.6$ in 
Fig.\ \ref{fig:B06_hdh_n12_3}. This energy corresponds to a
dip seen in Fig.\ \ref{fig:C_hdh}a. 
Due to the magnetic field and the scattering potential there is always
a scattering between these two channels irrespective of whether the
instate belongs to the $n=0$ (Fig.\ \ref{fig:B06_hdh_n12_3}a) or the $n=1$ 
(Fig.\ \ref{fig:B06_hdh_n12_3}b) mode. This is visible in the probability
density with interference pattern in all channels.

\begin{figure}[tbhp!]
      \includegraphics[width=0.45\textwidth]{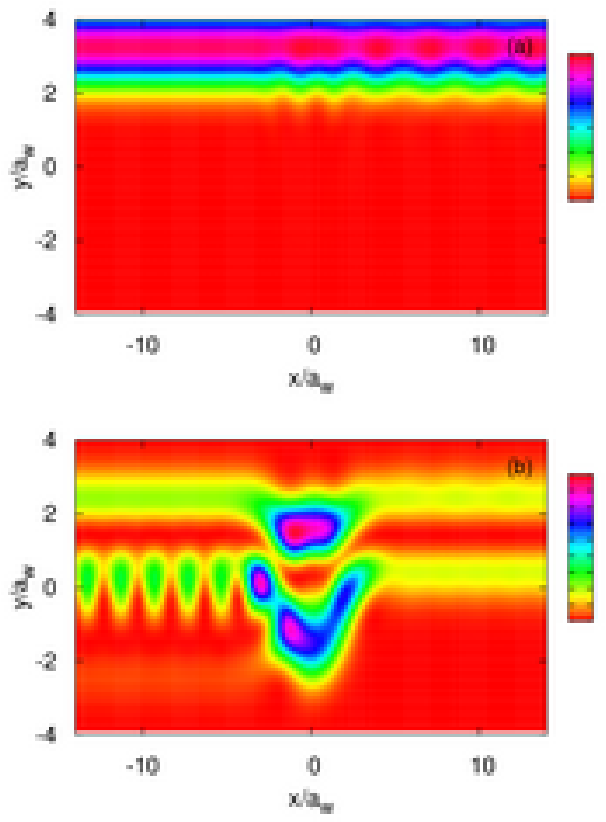}
      \caption{The probability density of the scattering state $\psi_{E}(x,y)$
               in the parabolic quantum wire in the presence of an
               embedded quantum dot (Fig.\ \ref{Ring_dot}), corresponding to
               the conductance in Fig.\ \ref{fig:C_hdh}b at $B=1.2$ T
               for a state with incident energy $X = 2.235$,               
               in mode $n=0$ (a), and $n=1$ (b).}
\label{fig:B12_hdh_n12_1}
\end{figure}

The situation is completely different at the higher magnetic field
$B=1.2$ T seen in Fig.\ \ref{fig:B12_hdh_n12_1}. Here the same scaled 
energy as before, $X = 2.235$, corresponds to a peak in 
the conductance displayed in Fig.\ \ref{fig:C_hdh}b.     
The value of the conductance peak indicates that there is very little
backscattering. The edge channel ($n=0$) is almost entirely tranmitted as
Fig.\ \ref{fig:B12_hdh_n12_1}a shows, but a quasi-bound state is seen in
Fig.\ \ref{fig:B12_hdh_n12_1}b belonging to the same subband as the instate. 

\begin{figure}[tbhp!]
      \includegraphics[width=0.45\textwidth]{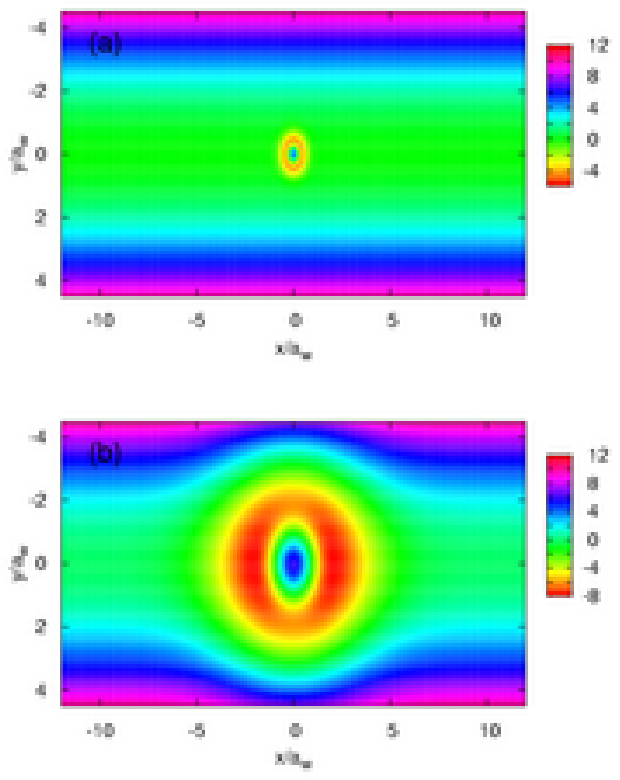}
      \caption{A contour plot of the potential of an embedded quantum ring in wire.
               (a) a ring with $\beta_1a_w^2=3.41$, and $\beta_2a_w^2=11.37$,
               corresponding to the ring in the upper panel of Fig.\ \ref{Ring_dot}.
               (b) a large ring with $\beta_1a_w^2=0.0682$(B=0), 
               and $\beta_2a_w^2(B=0)=0.682$.
               $E_0=\hbar\Omega_w=1.0$ meV, $a_w=33.7$ nm at $B=0$ T.}
\label{WireDot}
\end{figure}

The small quantum ring embedded in the broad wire (Fig.\ \ref{Ring_dot})
is to small to show any indication of Aharanov-Bohm oscillations, and as
the magnetic length gets smaller with increasing field strength the edge
states bypass the ring. To change this situation we also did calculations
for a larger ring shown in Fig.\ \ref{WireDot} compared to the smaller ring
used in the preceeding calculations. 

\begin{figure}[tbhp!]
      \includegraphics[width=0.45\textwidth]{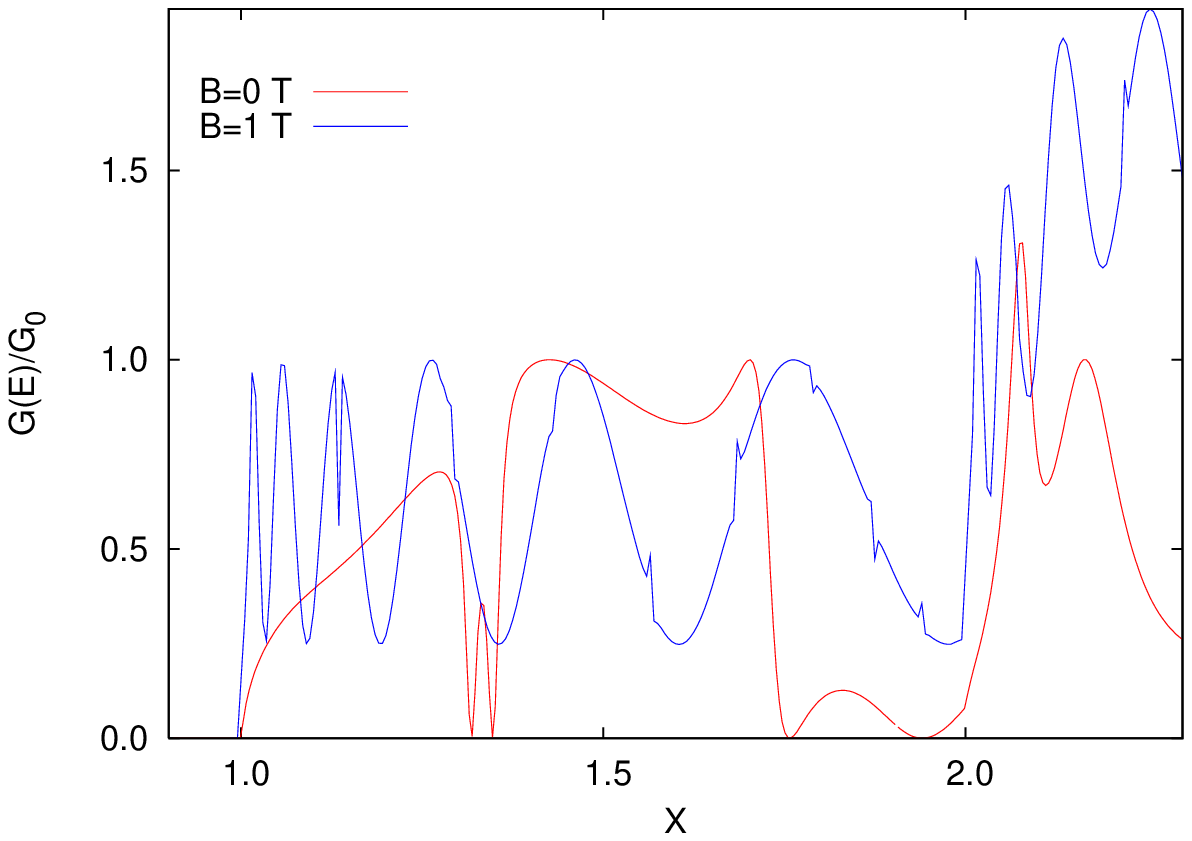}
      \caption{The conductance of a parabolic wire with a large embedded ring
               (corresponding to Fig.\ \ref{WireDot}b) in units of
               $G_0=2e^2/h$. $E_w=\hbar\Omega_w$, $V_1=-12$ meV, $V_2=18$ meV,
               $\hbar\Omega_0=1.0$ meV, $\beta_1 a_w^2(B=0)=0.0682$, 
               $\beta_2 a_w^2(B=0)=0.682$, and 13 subbands are included.}
\label{C_LH}
\end{figure}

Of course the parabolic confinement of the wire always leads to the situation
that at a high enough magnetic field the edge states will not be scattered by
the quantum ring potential, but now at an intermediate field strength the
magnetic length compares more favorably with the size scale of the ring
as can be seen in the conductance displayed in Fig.\ \ref{C_LH} for
both $B=0$ and 1 T. 
At $B=1$ T we see oscillations growing in wavelength with $E$ or $k_n(E)$
both for mode $n=0$ and 1.  

The oscillations in the conductance at $B=1$ T
are caused by a simple geometrical resonance where the wavelength of the 
scattering state in the ring has to compare appropriately with the circumference
of the ring to build constructive or destructive interference, i.\ e.\ a
Aharanov-Bohm like effect. This also explains the growing wavelength
of the oscillation with $(E-E_0)$. Even though the same condition
lies at the root of the energy spectrum of stationary states in a ring
in equilibrium we are not probing here the energy spectrum of the ring. 
We would like to 
mention that a similar oscillatory behavior of the conductance as a
function of the energy was reported by Sivan et al.\cite{Sivan89:1242} 
in the case of a quantum dot in high magnetic field. Since in high magnetic 
field the cyclotron radius is smaller than the ring radius one expects
the electrons to travel within the ring along skipping orbits before they
leave through the wire. 

\begin{figure}[tbhp!]
      \includegraphics[width=0.45\textwidth]{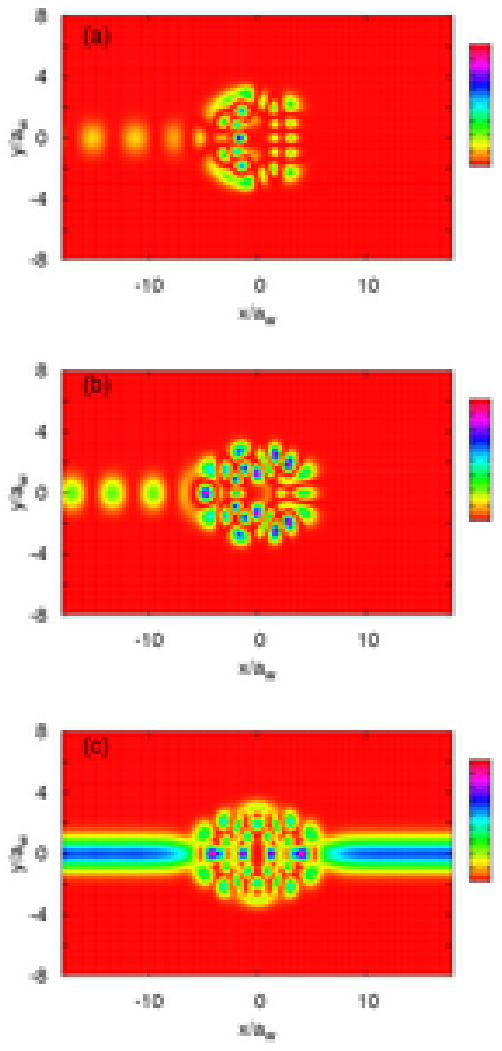}
      \caption{The probability density of the scattering state $\psi_{E}(x,y)$
               in the parabolic quantum wire in the presence of an
               embedded large quantum ring (Fig.\ \ref{WireDot}b), corresponding to
               the conductance in Fig.\ \ref{C_LH} at $B=0$ T
               for a state with incident energy $X = 1.319$               
               corresponding to a dip (a), $X = 1.347$ in a dip
               (b), and $X = 1.425$ at a maximum (c).
               The incoming mode is $n=0$.}
\label{SHB0_dens}
\end{figure}

The large size of the embedded ring in this case and it finite depth
mean that quasi-bound
states will not be of the same simple structure as seen for the smaller
ring. This can be verified by the probability densities shown in 
Fig.\ \ref{SHB0_dens} for the two dips at $X = 1.319$
and 1.347, and for the maximum at 1.425.
The total transmission of the only mode, $n=0$, in Fig.\ \ref{SHB0_dens}c
causes the perfect left right symmetry, but the probability density in
Fig.\ \ref{SHB0_dens}a corresponding to the dip at $X = 1.319$
reflects the asymmetry caused by the confining parabolic potential to
the ring seen in Fig.\ \ref{WireDot}b. The structure of the evanescent
states in Fig.\ \ref{SHB0_dens}a and b indicates that they are caused 
by states in the third and fifth energy bands and probably also states
in higher bands. This persistence of eigenstates 
or scarring of wave functions in open systems
has been discussed by Akis et at.\cite{Akis02:129} 
for quantum dots, and here we confirm it for an open quantum ring.

\begin{figure}[tbhp!]
      \includegraphics[width=0.45\textwidth]{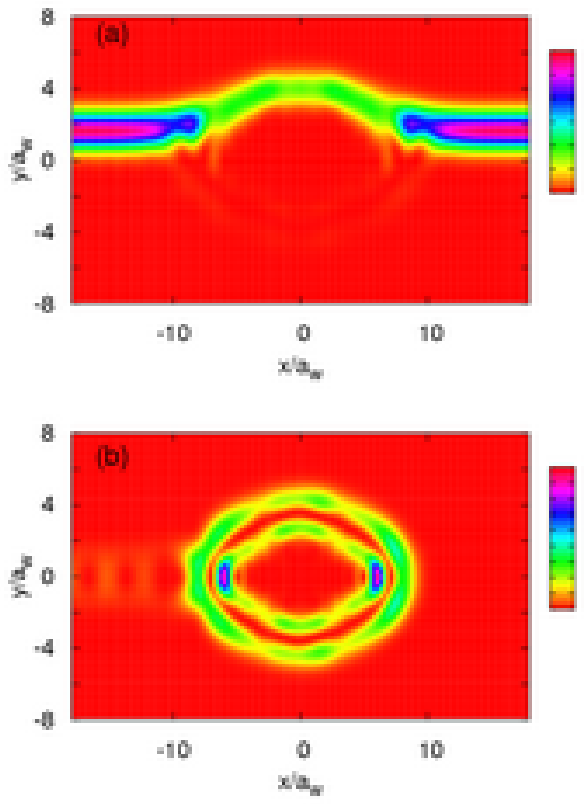}
      \caption{The probability density of the scattering state $\psi_{E}(x,y)$
               in the parabolic quantum wire in the presence of an
               embedded large quantum ring (Fig.\ \ref{WireDot}b), corresponding to
               the conductance in Fig.\ \ref{C_LH} at $B=1$ T
               for a state with incident energy $X = 1.46$               
               corresponding to a peak (a), $X = 1.135$ in a
               narrow dip (b). The incoming mode is $n=0$.}
\label{Wf_LHB}
\end{figure}

Superimposed on the Aharanov-Bohm like oscillations in the conductance in
Fig.\ \ref{C_LH} we have narrow resonances that are caused by interaction
with quasi-bound states of the ring. In Fig.\ \ref{Wf_LHB}a we show the
probability density for the scattering state for a Aharanov-Bohm peak
at $X = 1.46$. The density for a minimum in the oscillation
is similar except for the addition of the reflected wave. 
In Fig.\ \ref{Wf_LHB}b we display the probability density for the first
narrow resonance seen, at $X = 1.135$. Here we can identify 
a long lived evanescent state in the second subband. 
%

%
\section{Summary}

We have successfully extended a multiband transport formalism
build on the Lippmann-Schwinger equation in a magnetic field
to be able to describe an unbiased transport through a broad wire
with embedded small or large quantum dots and rings defined by a 
smooth potential. The calculation of the probability density for the
scattering states allows us to shed light on internal processes
and resonances that in some cases reflect interaction between
states in several subbands of the wire. We observe well known 
evanescent states and Fano resonances produced by these interactions. 
In the case of a large ring with finite width we observe Ahranov-Bohm 
type of oscillations superimposed with narrow resonances reflecting
its energy spectrum. 

Due to the wide range of system parameters used
we had to pay extra attention to the accuracy of the numerical methods
employed.        

%
\appendix
\section{\label{sec:appendix}Matrix elements of the scattering potential}

The matrix elements of the potential represented by a single Gaussian
function $V=V_0\exp{(-\beta_x x^2-\beta_y y^2)}$ is according to 
Eq.'s (\ref{Vpq}-\ref{Vp-qy})
\begin{equation}
      V_{nn'}(q,p) = V_0a_w\sqrt{\frac{\pi}{\beta_x}}
      \exp{\left[ -\frac{(q-p)^2}{4\beta_x} \right]}I_{nn'}(q,p) ,
\label{Vnnqp}
\end{equation}  
where
\begin{equation}
      I_{nn'}(q,p) = \int dy\: \phi_n^*(q,y)e^{-\beta_y y^2}\phi_{n'}(p,y).
\label{Inn}
\end{equation}  
Insertion of the expressions for the eigenfunctions (\ref{Phi_para}) yields
\begin{equation}
      \begin{split}
      I_{nn'}(q,p) =& \frac{\exp{[\frac{(s_n+s_{n'})^2}{4C}-\frac{s_n^2+s_{n'}^2}{2}}]}
      {2^{n+{n'}}\sqrt{C n! n'!}}\\
      &\sum_{p=0}^{n}\sum_{q=0}^{n'}\binom{n}{p}\binom{n'}{q}
      H_p(-\sqrt{2}s_n)H_p(-\sqrt{2}s_{n'})\\
      &\sum_{l=0}^{{\rm min}(n-p,n'-q)}2^ll!\binom{n-p}{l}
      \binom{n'-q}{l}\\ 
      &{\hspace*{2cm}}b^{\frac{N}{2}-l}
      H_{N-2l}\left( z\sqrt{\frac{2}{bC}}\right) ,   
      \end{split}
\label{Innn}
\end{equation}  
where $N = n+n'-p-q$, $s_n = y_0^{n} = k_na_w\omega_c/\Omega_w$, 
$z = (s_n+s_{n'})/(2\sqrt{C})$, $C=(1+\beta_ya_w^2)$, and 
$b = (1-2/C)$. When the variable $b$ assumes negative values
the combination $(\sqrt{b})^{N-2l}H_{N-2l}(\cdots /\sqrt{b})$ still
supplies the correct real value.


%
%
\begin{acknowledgments}
      The research was partly funded by the Research
      and Instruments Funds of the Icelandic State,
      the Research Fund of the University of Iceland, and the
      National Science Council of Taiwan.
      C.S.T. acknowledges the computational facility supported by the
      National Center for High-performance Computing of Taiwan.
      V.M. was supported by a NATO Science Fellowship.
\end{acknowledgments}

%
%
\bibliographystyle{apsrev}

%
%
%
\end{document}